\documentclass[11pt]{article}
\usepackage[dvips]{epsfig}

\advance \textwidth by -2in
\advance \textheight by -3in
\def\##1{\underline #1}
\def\=#1{\underline{\underline #1}}

\def\eps{\epsilon}
\def\epso{\epsilon_0}
\def\muo{\mu_0}
\def\ko{k_0}

\def\lambdao{\lambda_0}

\def\.{\mbox{ \tiny{$^\bullet$} }}

\def\epsa{\epsilon_{a}}
\def\epsb{\epsilon_{b}}
\def\epsc{\epsilon_{c}}
\def\epsdt{\tilde{\epsilon}_{d}}
\def\epsrefo{\=\eps_{\,ref}^o}

\def\ux{\#{u}_x}
\def\uy{\#{u}_y}
\def\uz{\#{u}_z}
\def\up{\#{u}_+}
\def\um{\#{u}_-}

\def\aal{a_L}
\def\aar{a_R}
\def\bbl{r_L}
\def\bbr{r_R}

\def\le{\left(}
\def\ri{\right)}
\def\les{\left[}
\def\ris{\right]}

\def\c#1{\cite{#1}}

\begin{document}
\vskip 0.4cm

\noindent {\large {\bf 
Post-- versus  pre--resonance characteristics of axially excited
chiral sculptured thin films}}
\vskip 0.4cm

\noindent  
{\bf Akhlesh Lakhtakia}\footnote{Tel: +1-814-863-4319, 
Fax: +1-814-863-7967, e-mail: AXL4@psu.edu}
 and {\bf Jason T. Moyer} 
\vskip 0.2cm
\noindent {\em CATMAS~---~Computational \& Theoretical Materials Sciences 
Group\\
Department of Engineering Science \& Mechanics\\
Pennsylvania State University, University Park, PA 16802--6812, USA}

\vskip 0.5cm

\noindent {\bf Abstract.} Axially excited chiral
sculptured thin films (STFs) are shown
to exhibit the circular Bragg phenomenon
in the pre--resonant (long--wavelength) regime
but not in some parts
of the post--resonant (short--wavelength) regime.
Chiral STFs act as very good polarization--independent
 reflectors in the vicinity of material resonances
in the latter regime. 

\vskip 0.2cm

\section{Introduction}

The chief optical signature of a chiral sculptured thin
film (STF) is the
circular Bragg phenomenon it displays on axial excitation \c{VLbook,WHL}.
A chiral STF is modeled as a
unidirectionally nonhomogeneous dielectric
continuum with a permittivity dyadic
that varies periodically along, say, the $z$ axis in a helicoidal
fashion.
Let  a   circularly polarized plane 
wave with  free--space wavelength $\lambdao$ axially
excite  a chiral STF of finite thickness. Provided 
that  the film thickness  is
sufficiently large and $\lambdao$ lies within the so--called Bragg regime,
the reflectance is much higher if the handedness of the incident
 plane wave matches its structural handedness than if otherwise.
Indeed,  light of one circular polarization~---~coincident
with the structural handedness of the chiral STF~---~effectively 
encounters a 
grating in the Bragg regime, while light of the opposite circular 
polarization
state does not \c{ML1,ML2}. The described
 phenomenon has been theoretically
established for chiral STF half--spaces as well \c{L2,LMc}.

Explicit in the theoretical research  \c{VLbook,
STF}, and implicit
in the  experimental research \c{SBB, HWAdv,HWC}, on STFs
thus far is the
supposition that the real parts of all cartesian components
of the permittivity dyadic of a chiral STF are  positive real.
Viewed through the lens of the Lorentz one--resonance model
\c{Kitt}, consideration has been essentially restricted
to the pre--resonant (or the long--wavelength) regime.
The response of a chiral STF in the post--resonant 
(or the short--wavelength) regime
is simply unknown, which lacuna motivated the work presented
here.

Accordingly, in this communication, we present the reflection 
characteristics
of an axially excited chiral STF half--space, and
compare the pre-- and the post--resonant signatures.
A note on notation:
Vectors are underlined, dyadics are double-underlined;
while an $\exp(-i\omega t)$ time-dependence is implicit,
with $\omega$ as the angular frequency.

\section{Theoretical Preliminaries}
 
The dielectric properties of the chiral STF are delineated by the
nonhomogeneous  permittivity dyadic \c{VLbook}
\begin{equation}
\=\eps(\#r) = \epso\, \=S_z(z,h)\.\=S_y(\chi)\.\epsrefo
\.\=S_y^{-1}(\chi)\.
\=S_z^{-1}(z,h)\, ; \qquad z \geq 0\, ,
\label{epsbasic}
\end{equation}
where
\begin{equation}
\label{epsref}
\epsrefo= \epsa \,\uz\uz +\epsb\,\ux\ux+\epsc\,\uy\uy\, .
\end{equation}
Here and hereafter, $\epso$ and
$\muo$ are the permittivity and
the permeability of free space (i.e., vacuum), respectively; 
$\ko =  2\pi/\lambdao
=\omega\,\sqrt{\muo\epso}$  is the wavelength in free space;
 $\ux$, $\uy$ and $\uz$
denote the unit cartesian vectors; while $\epsilon_{a,b,c}$
are complex--valued functions of $\omega$. The
rotation dyadic
\begin{eqnarray}
\nonumber
\=S_z(z,h) &=&
\uz\uz + \le \ux\ux+\uy\uy\ri\,\cos\frac{\pi z}{\Omega} \\ 
& & +\,\, h\,\le \uy\ux-\ux\uy\ri\,\sin\frac{\pi z}{\Omega}\, 
\end{eqnarray}
captures the helicoidal periodicity of the STF, with
 $2\Omega$ being the structural period. The integer $h=1$ for a
structurally right--handed chiral STF; and $h=-1$ for structural 
left--handedness.
 The tilt dyadic
\begin{equation}
\=S_y(\chi) = \uy\uy + (\ux\ux + \uz\uz) \, \cos\chi + (\uz\ux-\ux\uz)\,
\sin\chi
\end{equation}
represents the {\it locally\/} aciculate microstructure of the chiral STF.
In the
present context, the most interesting properties 
of axially excited chiral STFs
depend on $\Omega$, $\epsc$ and
\begin{equation}
\epsdt = \epsa\epsb/(\epsa \cos^2\chi+\epsb\sin^2\chi)\,.
\end{equation}

Suppose an arbitrarily polarized plane wave is
normally incident on the chiral STF
half--space from the lower half--space $z \leq 0$, which is vacuous.
As a result, a plane wave is reflected into the lower half--space.
The electric field phasor associated with the two plane
waves in the lower half--space is stated as 
\begin{eqnarray}
\nonumber
\#E(z) &=&  \le  \aal \, \up + \aar \, \um \ri \,
\exp\le i \ko z \ri\\
&+&\, \le  \bbl \, \um + \bbr \, \up \ri \,
\exp\le -i \ko z \ri \, ;  \qquad z \leq 0\, ,
\end{eqnarray}
and the corresponding magnetic field phasor
is then easily determined from the Faraday equation.
Here, the complex unit vectors $\#u_\pm = (\ux \pm i \uy)/\sqrt{2}$;
$\aal$ and $\aar$ are the known amplitudes
of the left-- and the right--circularly polarized 
components
of the incident plane wave; and
$\bbl$ and $\bbr$ are the unknown amplitudes 
of the reflected planewave components. 

Our attention is focussed in this communication on the reflection coefficients
entering the 2$\times$2 matrix in the following relation:
\begin{equation}
\label{eq15}
\les \begin{array}{cccc} \bbl \\ \bbr  \end{array}\ris  =
\les \begin{array}{cccc} r_{LL} & r_{LR} \\ r_{RL} & r_{RR}\end{array}\ris 
\,
\les \begin{array}{cccc} \aal \\ \aar  \end{array}\ris \, .
\end{equation}
These coefficients are doubly subscripted:
those with both subscripts identical refer to co--polarized,
while those with two different subscripts denote
cross--polarized, reflection. These coefficients
can be calculated by following the procedure
recorded elsewhere \c{L2,LMc}.

\section{Numerical Results and Conclusions}
Figure 1 shows plots of the reflectances
$R_{LL}=\vert r_{LL}\vert^2$, etc. as functions
of the parameter $\Omega/\lambdao$ when $h=1$, $\epsc = 3(1+0.01i) $
and $\epsdt = 3.3(1+0.02i)$. These plots hold for the pre--resonant
case, because ${\rm Re}\les \epsc\ris>0$ and ${\rm Re}\les \epsdt\ris>0$.
As predicted, the circular Bragg phenomenon is in evidence as
the peak in the plot of $R_{RR}$ at $\Omega/\lambdao \approx 0.284$.
In contrast, $R_{LL}\simeq 0$ for all values of
$\Omega/\lambdao\in \les 0.1,0.5\ris$. The cross--polarized
reflectances are quite small and
also virtually indistinguishable from each other.

In the very short--wavelength portion of the post--resonant regime, 
${\rm Re}\les \epsc\ris>0$ and ${\rm Re}\les \epsdt\ris>0$ as well. The
circular Bragg phenomenon will be exhibited in that portion
of the electromagnetic spectrum,
just as in the pre--resonant regime, provided that
continuum electromagnetic theory remains valid.

However,
closer to the material resonances, ${\rm Re}\les \epsc\ris<0$ 
and ${\rm Re}\les \epsdt\ris<0$ in the post--resonant regime.
Figure 2 shows plots of the same reflectances as in
Figure 1, except
that $\epsc = 3(-1+0.01i) $
and $\epsdt = 3.3(-1+0.02i)$. 
The absence of the circular Bragg phenomenon is conspicuous.
Indeed, while both co--polarized reflectances are almost negligible,
both cross--polarized reflectances are extremely high. These
plots indicate that a post--resonant 
 chiral STF will reflect very well in the vicinity
of the material resonances,
in the same way that metals do.

The inescapable conclusion is that the circular Bragg
phenomenon can be exhibited  in the pre--resonance regime
but not throughout the post--resonant regime. Thus, soft ultra--violet
reflectance spectrums of chiral STFs are unlikely
to be interesting for circular--polarization--sensitive reflection
applications.
Our results also suggest the futility of transmission measurements
for characterizing chiral STFs
in the long--wavelength portion of the post--resonant regime.

Because of mathematical
isomorphism, similar conclusions should hold~---~other 
factors aside~---~for chiral liquid crystals too \c{Chan}.
Furthermore, as materials with negative real permittivity
in the microwave regime have now become artificially possible
\c{PHRS,SSS,LSM}, our conclusions should also apply for
synthetic laminar cholesteric materials for duty in that portion
of the electromagnetic spectrum.

\newpage
\begin{figure}[!ht]
\centering \psfull
\epsfig{file=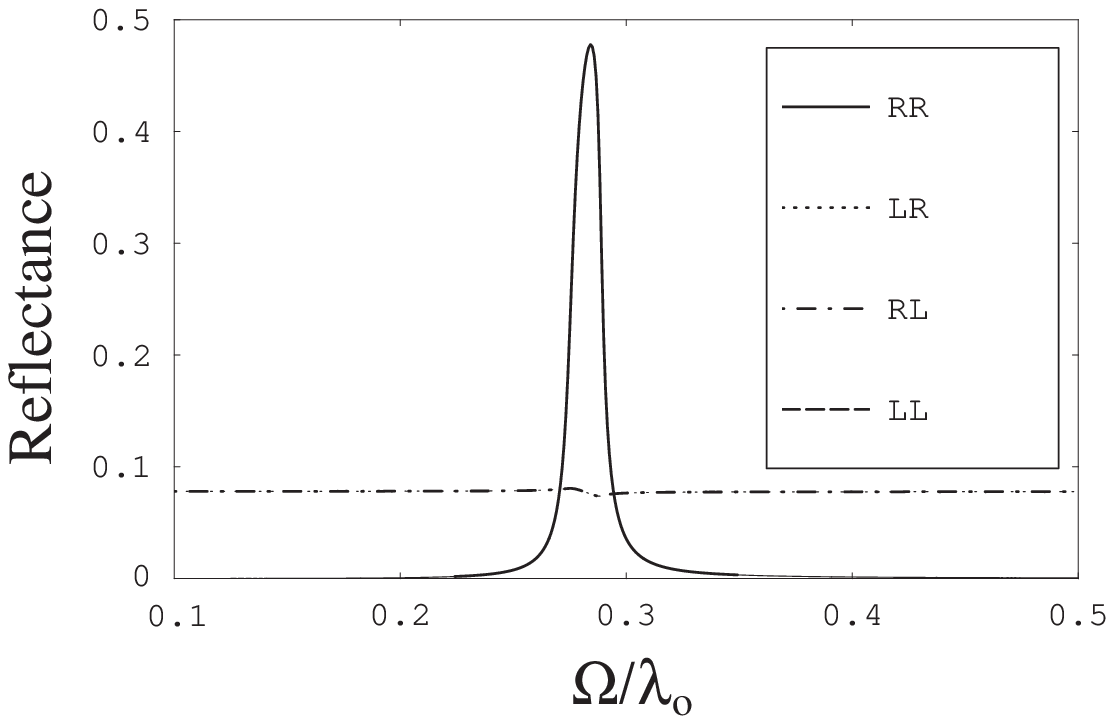}
\bigskip\bigskip
\caption{Computed values
of the  reflectances 
as functions of $\Omega/\lambdao$,
when $h=1$, $\epsc = 3(1+0.01i) $
and $\epsdt = 3.3(1+0.02i)$. Note
that $R_{RL} \simeq R_{LR}$, as pointed
out elsewhere \c{LMc}; while $R_{LL} \simeq 0$
cannot be discerned at the scale of the graph.}
\end{figure}


\newpage
\begin{figure}[!ht]
\centering \psfull
\epsfig{file=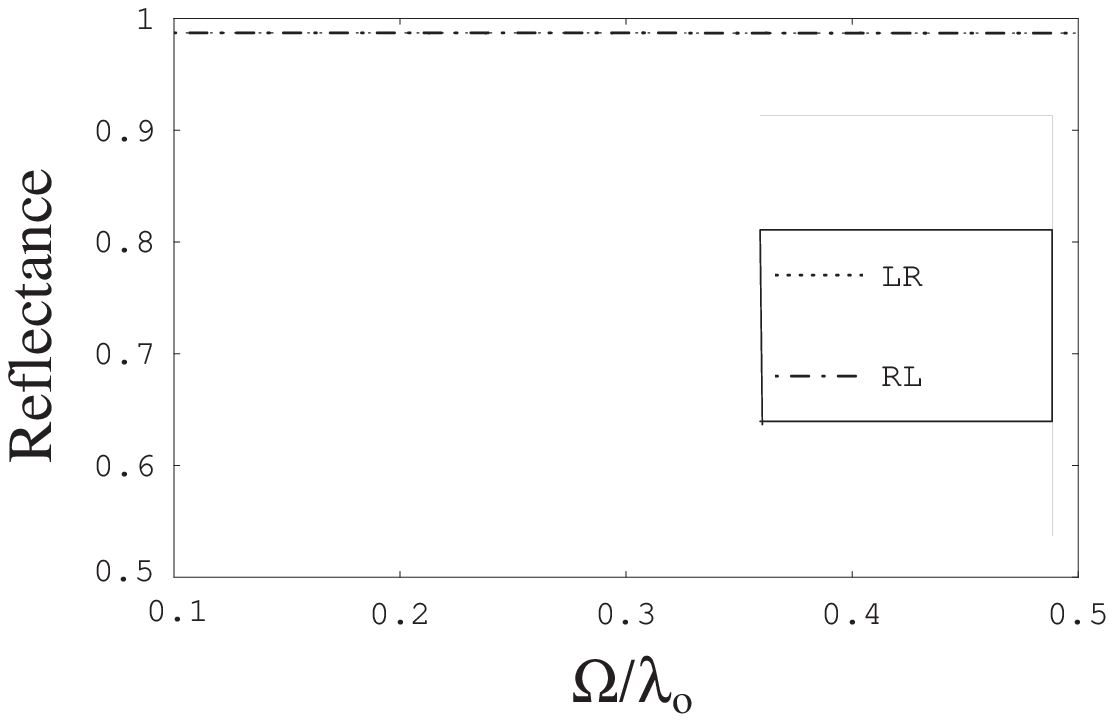}
\bigskip\bigskip
\caption{Computed values
of the  cross--polarized reflectances 
as functions of $\Omega/\lambdao$,
when $h=1$, $\epsc = 3(-1+0.01i) $
and $\epsdt = 3.3(-1+0.02i)$. Note
that $R_{RL} \simeq R_{LR}$; while $R_{RR} \simeq 0$ 
and $R_{LL}\simeq 0$ cannot be discerned at the scale of the graph
and are therefore not shown.}
\end{figure}



\newpage
\begin{thebibliography}{99}

\bibitem{VLbook}
Venugopal VC, Lakhtakia A:
Sculptured thin films: Conception,
optical properties, and applications. In: Singh ON,
Lakhtakia A (eds):
Electromagnetic Fields in Unconventional Materials and Structures.
Chap. 5.
Wiley, New York, 2000 

\bibitem{WHL}
Wu Q, Hodgkinson IJ, Lakhtakia A:
Circular polarization filters made of chiral sculptured
thin films: experimental and simulation results.
Opt. Engg. {\bf 39} (2000) 1863--1868


\bibitem{ML1}
McCall MW, Lakhtakia A:
Development and assessment of coupled wave
theory of axial propagation in thin--film helicoidal
bianisotropic media. Part 1: reflectances and transmittances.
J. Modern Opt. {\bf 47} (2000) 973--991

\bibitem{ML2}
McCall MW, Lakhtakia A:
Development and assessment of coupled wave
theory of axial propagation in thin--film helicoidal
bi--anisotropic media. Part 2: dichroisms, ellipticity
transformation and optical rotation.
J. Modern Opt. {\bf 48} (2001) 143--158

\bibitem{L2}
Lakhtakia A: On percolation and circular Bragg phenomenon
in metallic, helicoidally periodic, sculptured thin films.
Microw. Opt. Technol. Lett. {\bf 24} (2000) 239--244 [Two
typographical errors need correction: (i) $a_4\equiv 0$ (but not $a_3
\equiv 0$)
when $P_{4z}/\vert a_4\vert^2 < 0$; and $r_L-r_R$ should be replaced
by $r_L+r_R$ in Eq. (17).]

\bibitem{LMc}
Lakhtakia A, McCall MW: Simple expressions for Bragg reflection
from axially excited chiral sculptured thin films. J. Modern
Opt. {\bf 00} (2002) 000--000 (accepted in October 2001 for publication)

\bibitem{STF}
Lakhtakia A, Messier R: Sculptured thin
films. OSA Opt. Photon. News {\bf 12} (2001)
26--31 (September issue)

\bibitem{SBB}
Sit JC, Broer DJ, Brett MJ: Liquid
crystal alignment and switching in porous chiral optical
materials. Adv. Mater. {\bf 12} (2000) 371--373


\bibitem{HWAdv}
Hodgkinson I, Wu Qh: Inorganic chiral optical
materials. Adv. Mater. {\bf 13} (2001) 889--897

\bibitem{HWC}
Hodgkinson I, Wu QH, Collett S: Dispersion equations
for vacuum--deposited tilted--columnar biaxial media.
Appl. Opt. {\bf 40} (2001) 452--457

\bibitem{Kitt}
Kittel C: Introduction to Solid State Physics, Chap.
13. Wiley Eastern, New Delhi, India 1974

\bibitem{Chan}
Chandrasekhar S: Liquid Crystals. Cambridge
University Press, Cambridge, UK 1992

\bibitem{PHRS}
Pendry JB, Holden AJ, Robbins DJ, Stewart WJ:
Low frequency plasmons in thin--wire structures.
J. Phys.: Condens. Matter {\bf 10} (1998) 4785--4809

\bibitem{SSS}
Shelby RA, Smith DR, Schultz S: Experimental
verification of a negative index of refraction.
Science {\bf 292} (2001) 77--79

\bibitem{LSM}
Lakhtakia A, Slepyan GY, Maksimenko SA:
Towards cholesteric absorbers for microwave frequencies.
Int. J. Infrared Millim. Waves {\bf 22} (2001)
999--1007

\end{thebibliography}
\end{document}